\documentclass[%
 aip,
 amsmath,amssymb,
 reprint,%
]{revtex4-1}

\usepackage{graphicx}
\usepackage{dcolumn}
\usepackage{bm}

\usepackage[utf8]{inputenc}
\usepackage[T1]{fontenc}
\usepackage{mathptmx}

\usepackage{amsmath}
\usepackage{amsthm}
\usepackage[colorlinks=true,urlcolor=blue,citecolor=blue,linkcolor=blue]{hyperref}
\usepackage[shortlabels]{enumitem}
\usepackage{amsfonts}
\usepackage{graphicx}
\usepackage{bbold}
\usepackage{color}
\usepackage{hyperref}
\usepackage{verbatim}
\usepackage{cases}
\usepackage{amsfonts}
\usepackage{amssymb}
\usepackage[]{mathrsfs}
\usepackage{xcolor}
\usepackage{epsfig}
\usepackage{bm}
\usepackage{setspace}
\usepackage{enumerate}
\usepackage{dcolumn}
\usepackage{bm}
\usepackage{enumitem}

\usepackage{graphicx}
\usepackage{dcolumn}
\usepackage{bm}
\usepackage[FIGBOTCAP,small,tight]{subfigure}
\usepackage{color}

\newcommand{\bra}[1]{\left< #1 \right\vert}
\newcommand{\ket}[1]{\left\vert #1 \right>}

\newcommand{\llav}[1]{\left\lbrace #1 \right\rbrace}

\begin{document}

\preprint{AIP/123-QED}

\title[]{\textcolor{blue}{Photochemical dynamics under incoherent illumination: light harvesting in self-assembled molecular J-aggregates}\vspace{1mm}}

\author{Luis Felipe Morales-Curiel}
\affiliation{Instituto de Ciencias Nucleares, Universidad Nacional Aut\'onoma de M\'exico, Apartado Postal 70-543, 04510 Cd. Mx., M\'exico}


\author{Roberto de J. Le\'{o}n-Montiel}%
\email{roberto.leon@nucleares.unam.mx}
\affiliation{Instituto de Ciencias Nucleares, Universidad Nacional Aut\'onoma de M\'exico, Apartado Postal 70-543, 04510 Cd. Mx., M\'exico}

\date{\today}

\begin{abstract}
Transport phenomena in organic, self-assembled molecular J-aggregates have long attracted a great deal of attention due to its potential role in designing novel organic photovoltaic devices. A large number of theoretical and experimental studies have been carried out describing excitonic energy transfer in J-aggregates under the assumption that excitons are induced by a coherent laser-light source or initialized by a localized excitation on a particular chromophore. However, these assumptions may not provide an accurate description to assess the efficiency of J-aggregates, particularly as building blocks of organic solar cells. In natural conditions, J-aggregates would be subjected to an incoherent source of light (as is sunlight), which would illuminate the whole photosynthetic complex rather than a single molecule. In this work, we present the first study of the efficiency of photosynthetic energy transport in self-assembled molecular aggregates under incoherent sunlight illumination. By making use of a minimalistic model of a cyanine dye J-aggregate, we demonstrate that long-range transport efficiency is enhanced when exciting the aggregate with incoherent light. Our results thus support the conclusion that J-aggregates are indeed excellent candidates for devices where efficient long-range incoherently-induced exciton transport is desired, such as in highly efficient organic solar cells.
\end{abstract}

\maketitle

\section{Introduction}

In recent years, there has been a great excitement due to the experimental observation of long-lived electronic coherences in bacterial and algal light-harvesting complexes. \cite{engel_2007,engel_2010,scholes_2010,Wong:2012jd} Since their observation, it has been argued that coherence effects might play an important role in the efficient energy transport of photosynthetic systems. \cite{ishizaki_2009,Ishizaki:2009ky,whaley_2010,alexandra_2010,schulten2012,Fujita:2012uo,Rob2} However, some of these effects have been controversial, as it has been argued that sunlight, being an incoherent source of photons, does not sustain transient coherent phenomena,\cite{zimanyi_2010,Brumer:2011ty,miller2012,kassal2013,han2013,Brumer2018-1,Brumer2018-2} and that the efficiency can be understood as a consequence of another type of coherent phenomena that is inherent to the light-harvesting material, such as spatial delocalization of excited electronic states, which survives even under incoherent illumination, and may increase energy transport efficiency. \cite{roberto2014_jpcb,fassioli2014}

A novel design strategy for artificial light-harvesting systems relies on the use of organic molecular aggregates that possess coherent exciton properties, such as cyanine dye J-aggregates.\cite{martinez2010,chen2013} Discovered almost 80 years ago, \cite{jelley1936,scheibe1936} J-aggregates are chemical structures formed by organic fluorescent dye molecules, which can be identified by a narrow absorption band (J-band) that is shifted to a longer wavelength with respect to the monomer absorption band.\cite{wurthner2011,higgins1996} These aggregates are characterized by unique properties including a large absorption cross section, small Stokes shift of the fluorescence line, and efficient energy transfer within the aggregate. \cite{wurthner2011} Because of these properties, dye aggregates have led to various technological applications, for instance, J-aggregates have been used as photography \cite{james_book} and solar-cell \cite{Saito1999,Sayama2002} sensitizer dyes, as fluorescent markers of mitochondrial membrane potentials in living cells, \cite{smiley1991} and as enablers of efficient energy transfer in quantum dot/J-aggregate blended films. \cite{walker2010} More recently, J-aggregates have been recognized as a promising material for the development of efficient, low-cost artificial light-harvesting devices. \cite{saikin2013,caram2016}

Although much work has been devoted in describing the exciton dynamics of self-assembled molecular aggregates, most of the studies assume that excitation is initially localized on a particular chromophore. \cite{valleau2012,hestand2015,saikin2017,hestand2018} This assumption might be incorrect: unless the system is so disordered that each eigenstate is effectively localized on one site, the initial state will be at least partially delocalized. In this work, we examine energy transport in molecular J-aggregates in a more realistic scenario, namely under incoherent delocalized illumination. Remarkably, we find that long-range transport efficiency is enhanced when exciting the molecular aggregate with incoherent light. Our results thus place J-aggregates as excellent candidates for devices where efficient long-range incoherently-induced exciton transport is sought, such as in efficient organic solar cells.

\section{Theoretical Methods}

Following previous authors, \cite{valleau2012} we consider the dynamics of excitons in a 2D monolayer J-aggregate comprising $N$ fluorescent cyanine dye molecules [see Figs. 1(a,b)]. This system, in the single-exciton basis, can be described by a tight-binding Hamiltonian of the form \cite{May:2011vy}

\begin{equation}
\hat{H} =  \sum_{i = 1}^{N} \epsilon_{i} \ket{i}\bra{i} + \sum_{i \neq j}^{N} J_{ij}\ket{i}\bra{j},
\end{equation}

\noindent where $\epsilon_{i}$ and $J_{ij}$ stand for the excitation energy of the $i$th-site energy and the coupling between sites $i$ and $j$, respectively. We will denote the eigenstates of $H$ as $\ket{e_i}$, with energies $E_{i}$, that is,

\begin{equation}
H\ket{e_{i}} = E_{i}\ket{e_i}.
\end{equation}

We will also denote the state where no excitons are present in the aggregate as $\ket{g}$, and the state where the exciton has been transferred to a sink, which will be used to measure the transport efficiency at different lengths of the aggregate, as $\ket{\text{sink}}$. Note that neither $\ket{g}$ nor $\ket{\text{sink}}$ are coupled to the states $\ket{i}$ through $H$.

The exciton dynamics in a photosynthetic system interacting with its environment is in general complicated and non-Markovian. \cite{chen2011} However, because the scope of this work is to provide qualitative features of exciton transport in J-aggregates under incoherent illumination, we will use a Markovian model that, although simplistic, captures the essential physics, \cite{haken1972,haken1973} and explains important, experimentally-observed features of excitons in molecular J-aggregates. \cite{eisele2012,moix2013,chuang2016,dijkstra2016,dijkstra2016,kriete2019} We then assume that the system-bath interaction can be modeled using a Lindblad master equation, for the system density matrix $\rho$, of the form \cite{book_open}

\begin{equation}
\frac{\partial \rho}{\partial t} = -\frac{i}{\hbar}[H,\rho] + L_{\text{deph}}[\rho] + L_{\text{diss}}[\rho] + L_{\text{sink}}[\rho],
\end{equation}

\noindent where
\begin{equation}
 L_{\text{deph}}[\rho] = \sum_{i = 1}^{N} 2\gamma_{i}(\ket{i}\bra{i}\rho\ket{i}\bra{i} - \frac{1}{2} \{\ket{i}\bra{i},\rho\}),
\end{equation}

\noindent describes a pure dephasing process that attenuates the coherence between different sites at a rate $\gamma_{i}$. Note that the symbol $\{.,.\}$ stands for the anticommutator. The third term in the Lindblad equation [Eq. (3)] describes a mechanism in which the excitation can be lost to the environment, this means that the exciton can be recombined at each site at rate $\Gamma_{i}$:

\begin{equation}
L_{\text{diss}}[\rho] = \sum_{i = 1}^{N} 2\Gamma_{i}(\ket{g}\bra{i}\rho\ket{i}\bra{g} - \frac{1}{2} \{\ket{i}\bra{i},\rho\}).
\end{equation}
However, the exciton lifetime in J-aggregates is usually longer $(\sim \text{ps})$ than the relevant transport phenomena, which occurs on a shorter time scale $(\sim \text{fs})$.  \cite{valleau2012,Jlibro} Therefore, in our simulations, the dissipation rate is assumed to be zero ($\Gamma_{i} = 0$). Finally the last term in Eq. (3) represents the process where the exciton is transferred to a sink at a rate $\Gamma_{sink}$:
\begin{equation}
L_{\text{sink}}[\rho] = 2\Gamma_{sink}(\ket{\text{sink}}\bra{k}\rho\ket{k}\bra{\text{sink}} - \frac{1}{2} \{\ket{k}\bra{k},\rho\}),
\end{equation}

\noindent where the site $k$ is the chromophore from which energy is irreversibly lost to the sink site. Because we are interested in describing the transport efficiency of the aggregate, we define the efficiency $\eta$ as the probability that the energy will arrive at the sink \cite{roberto2014_jpcb}

\begin{equation}
\eta = \lim_{t \to \infty } \bra{\text{sink}}\rho(t)\ket{\text{sink}}.
\end{equation}

As mentioned above, the typical way to treat excitations in the study of photosynthetic light-harvesting complexes is to assume that the excitation is initially localized on a single site. \cite{valleau2012,hestand2015,saikin2017,hestand2018,Ishizaki:2009ky,Fujita:2012uo,Alan2,Plenio,Chin,Kassal,Caruso} This type of excitation is described by
\begin{equation}
\rho_{\text{local}} = \ket{\text{ini}}\bra{\text{ini}},
\end{equation}
with $\ket{\text{ini}}$ describing the site where the excitation is initialized, which we assume to be located in the central site of the J-aggregate. \cite{valleau2012} Note that this type of excitation might not resemble what happens in real conditions. Firstly, the size of the cyanine molecules is so small ($\sim 9$ nm)\cite{valleau2012} compared to the absorbing wavelengths ($\sim 475$ nm) that excitation of the molecular aggregate cannot be thought of as localized in a single site, but as an excitation that is delocalized amongst all the eigenstates of the molecular complex. \cite{May:2011vy,roberto2014_jpcb}

Secondly, in natural conditions, excitation of photosynthetic complexes is not impulsive, except in controlled ultrafast experiments. Rather, excitation occurs through a steady-state, \cite{kassal2013,chenu2013} where an external energy source continuously pumps the system and excitation energy is continuously lost to either the environment or a sink. In this situation, the master equation of the system dynamics includes an additional term that transfers population, with a rate $\Gamma_{\text{pump}}$, from the ground state to the eigenstates, \cite{tscherbul2014,tscherbul2018}
\begin{equation}
L_{\text{pump}}[\rho] = 2\Gamma_{\text{pump}}\llav{\sum_{i=1}^{N}P(e_{i})\ket{e_{i}}\ket{g}\rho\bra{g}\bra{e_{i}} - \frac{1}{2}\llav{\ket{g}\bra{g},\rho}},
\end{equation}
where $P(e_{i})$ describes the probability of populating each of the eigenstates. It is worth pointing out that, in the steady-state context, the natural definition of efficiency is the ratio of exciton (or energy) flux to the sink from the incoming flux that pumps the system; consequently Eq. (7) may seem inappropriate. However, Jesenko and \v{Z}nidari\v{c} have shown that the efficiency of a steady-state process is equal to the efficiency of the corresponding impulsive case; \cite{jesenko2013} and thus we can keep using Eq. (7) to describe the transport efficiency for what is in reality a steady-state process.

Thirdly, in nature, photosynthetic systems are excited by incoherent sunlight. In this situation, the various frequencies of the light only excite populations of the eigenstates of the molecular complex, with which they are resonant and not coherences between them. \cite{roberto2014_jpcb,Jiang:1991fk,Mancal:2010kc,Brumer:2011ty} If we assume that each transition has the same oscillator strength, i.e. the probability of exciting the eigenstates is the same for all of them, incoherent light will create a mixture of eigenstates in proportion to the intensity of the light spectrum at the transition frequency, and the initial state of the system would thus be given by

\begin{equation}
\rho_{\text{incoh}} = \frac{1}{\mathcal{N}} \sum_{i=1}^{N} S(E_{i})\ket{e_{i}}\bra{e_{i}},
\end{equation}
where $S(E_{i})$ is the frequency spectrum of the light, which we assume to be the same as that of the J-aggregate absorption spectrum [see Fig. 1(c)], and $\mathcal{N}$ is a normalization constant.

\begin{center}
\begin{figure*}[t!]
    \includegraphics[width=17.75cm]{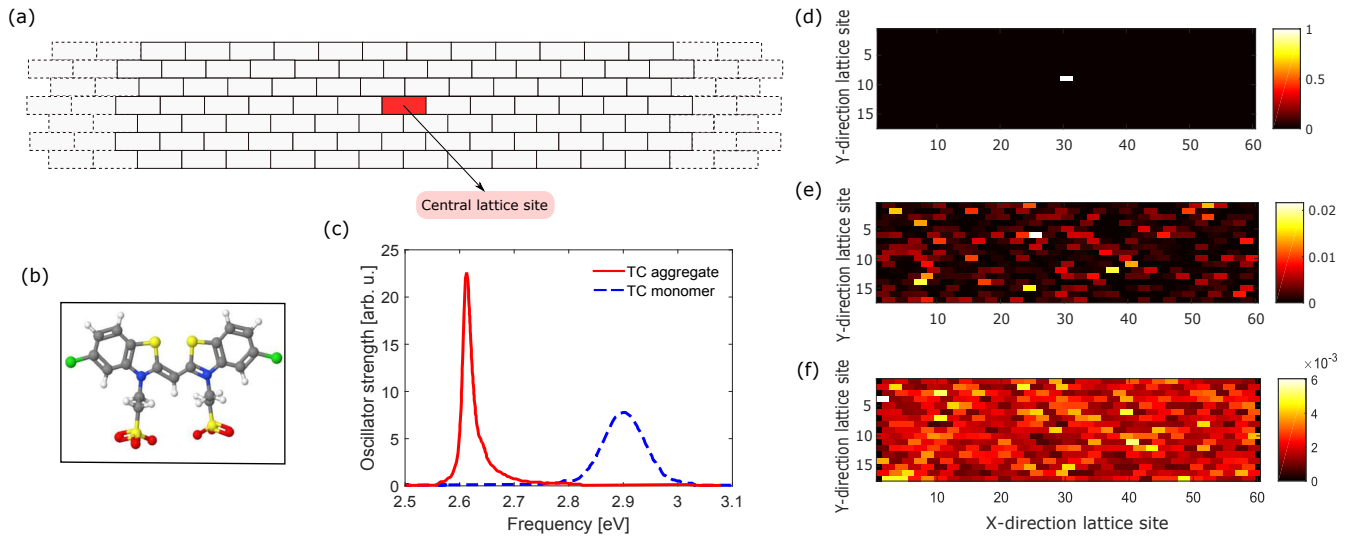}
       \caption{(a) Schematic representation of the 2D molecular aggregate comprising  five hundred 5,5'dichloro-3,3'-disulfopropyl thiacyanine (TC) monomers in a brickstone configuration. (b) Three-dimensional model of a TC monomer. (c) Absorption spectra of the TC aggregate (red solid line) and the TC monomer (blue dashed line). Note the red shift of the aggregate's spectrum with respect to the monomer absorption band. The absorption-spectrum data was taken from Ref. \cite{valleau2012}. (d), (e) and (f) show the initial energy distribution for the three different types of illumination considered here, namely localized, coherently delocalized and incoherently delocalized excitation, respectively.}
    \label{fig:figure1}
\end{figure*}
\end{center}

For the sake of completeness, we can also consider a coherently delocalized excitation, \cite{roberto2015} which can be produced by a coherent source, such as a laser pulse. In this case, a superposition, rather than a mixture, of excited states is induced. \cite{roberto2014_jpcb,Jiang:1991fk,Mancal:2010kc,Brumer:2011ty} The state will depend on the properties of the pulse, such as its spectrum, peak power, and duration; however, to enable comparison with Eq. (10), we can assume the same spectrum and write the initial state as
\begin{equation}
\rho_{\text{coh}} = \frac{1}{\mathcal{N}} \sum_{i=1}^{N} \sum_{j=1}^{N} \sqrt{S(E_{i})S(E_{j})} \ket{e_{i}}\bra{e_{j}}.
\end{equation}
Note that Eq. (11) is a particular example of coherent excitation with light bearing the same spectrum as the state in Eq. (10). A more general coherent state would include control over the phases of the off-diagonal terms.

\section{Results and Discussion}

In the results that follow, we have considered a 2D monolayer J-aggregate comprising five hundred 5,5'dichloro-3,3'-disulfopropyl thiacyanine (TC) molecules arranged in a brickstone lattice, as depicted in Fig. 1(a). The site energies $\epsilon_{i} = 3.3\;\text{eV}$ are taken from Reference \cite{valleau2012} whereas the nearest-neighbor couplings $J_{ij}$ are calculated using the extended dipole model, \cite{czikklely1970} where the interaction coefficient is obtained as the sum of Coulomb interactions between the transition charges located in different molecules. Interestingly, the topology of the J-aggregate allows us to simplify the calculation and define two characteristic couplings given by $J^{(H)} =  -0.1364\;\text{eV} $ and $J^{(V)} = -0.1077\;\text{eV}$, which represent the interaction of one molecule with its neighbor in the horizontal (left and right) and vertical (above and below) direction, respectively.

For the sake of simplicity, all sites are assumed to be in identical local environments. This allows us to set the dephasing rates equal for all sites. In the Markovian limit, these rates are given by \cite{Alan2,valleau2012} $\gamma_{i} \approx k_{B}T$, with $k_{B}$ being the Boltzmann constant and $T$ the temperature of the aggregate's environment. This helps us to find that, at room temperature, the dephasing rates are $\gamma_{i} \approx 26\;\text{meV}$. Finally, we take the transfer rate from a monomer in the aggregate to the sink as the optimum value for all initial conditions, namely localized, coherent and incoherent excitations. By testing different rates at different sink positions we have found that $\Gamma_{\text{sink}} = 0.5\;\text{fs}^{-1}$ gives the best transfer efficiency in all cases. Note that this value allows us to make a fair comparison between the three initial excitation conditions.

We have calculated the efficiency $\eta$ of excitonic transport using the three different initial conditions given above [see Figs. 1(d-f)]. The efficiency is measured by placing a sink that interacts with a single molecule located at the right-hand (or left-hand) side of the central site along the horizontal direction. The time evolution was set to 40 fs to avoid border effects. Figure 2 shows the transport efficiency for increasingly larger distances (in the horizontal direction) from the central site of the lattice. Note that, as expected, at short distances the best efficiency is obtained for the initially localized excitation. However, for the relevant case of long-range exciton transport, the incoherently delocalized excitation exhibits the best efficiency. This is due to the fact that the absence of off-diagonal elements in the incoherent initial state (written in the eigenstate basis) creates a highly delocalized state in the site basis [see Fig. 1(f)], which ultimately leads to an enhanced long-range transport efficiency. We have quantified the efficiency enhancement in the long-range regime (121.5 nm from the central network site) and found that incoherent illumination shows a $27\%$ higher efficiency than a coherent initial excitation condition, whereas compared to the localized excitation the incoherent illumination shows a $93\%$ enhanced efficiency.

Interestingly, by carefully analyzing the dephasing rate used in Eq. (4), $\gamma_{i} \approx 26\;\text{meV}$, one can find that the dynamics of excitons under these conditions is rather coherent. This is why even an initially localized excitation can reach a far-lying sink. However, in a hypothetical case of strong dephasing, an initially localized excitation would be likely to stay in its initial position due to the Zeno effect. \cite{Alan2} This would lead to an extremely low transport efficiency for the localized excitation condition; thus implying that, even in a large-dephasing scenario, the incoherently delocalized excitation condition is the best option for achieving a higher long-range transport efficiency.

From the results presented above, we can clearly see that the most realistic initial condition, namely incoherently delocalized excitation, delivers the highest energy transport efficiency. It is important to remark that, in general, the sink (or reaction center) is not a localized site, but a complex that interacts weakly with the light-harvesting molecular aggregate. In this situation, energy transport takes place via the eigenstates of both complexes, \cite{May:2011vy} and the efficiency becomes maximum for all three types of excitation. \cite{roberto2014_jpcb} Therefore, our results show that in the presence of a localized sink, the best long-range transport efficiency is found under incoherent illumination.

\begin{figure}[t!]
\includegraphics[width=8.5cm]{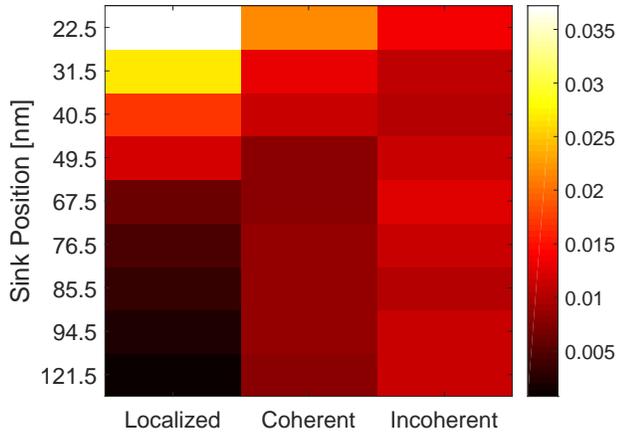}
\caption{Transport efficiency as a function of the separation between the sink and the central site of the molecular lattice. The first column shows the efficiency for the localized excitation, whereas the second and third columns show the efficiency for coherently and incoherently delocalized excitation, respectively. The sink's transfer rate, $\Gamma_{\text{sink}} = 0.5\;\text{fs}^{-1}$, is the optimum for all initial conditions, and the time evolution is set to 40 fs.}
\centering
\end{figure}

\section{Conclusion}

In this work, we have analyzed the short- and long-range transport efficiency of a 2D monolayer J-aggregate under different illumination conditions. By making use of a simple model we have shown that while short-range transport efficiency is the highest in the localized excitation case, the more physically relevant long-range energy-transport regime is enhanced (by 27\% and 93\% when compared to the coherent and localized excitation, respectively) when exciting the molecular aggregate with incoherent light. Our results thus predict that J-aggregates will exhibit their best performance under incoherent illumination, and support the conventional wisdom that J-aggregates are excellent candidates for designing devices where efficient long-range incoherently-induced exciton transport is desired, such as in highly efficient organic solar cells.


\begin{acknowledgments}
This work was supported by CONACyT under the project CB-2016-01/284372.
\end{acknowledgments}

\bibliography{BIB}

\end{document}